\documentclass{article}
\usepackage[utf8]{inputenc}
\usepackage{amsmath}
\usepackage[margin=1in]{geometry}
\usepackage{cite}
\usepackage{bm}
\usepackage{graphicx, calc}
\usepackage{caption}
\usepackage{amsfonts}
\usepackage{mathtools}
\usepackage{longtable}
\usepackage{array}
\usepackage{authblk}
\usepackage{orcidlink}
\usepackage{setspace}
\DeclarePairedDelimiter{\ceil}{\lceil}{\rceil}

\title{Evaluating The Impact Of Species Specialisation On Ecological Network Robustness Using Analytic Methods}
\author[1]{Chris Jones \orcidlink{0000-0003-4698-5573}}
\author[2]{Damaris Zurell \orcidlink{0000-0002-4628-3558}}
\author[3]{Karoline Wiesner \orcidlink{0000-0003-2944-1988}}
\affil[1]{School of Mathematics, University of Bristol, Bristol, UK}
\affil[2]{Institute of Biochemistry and Biology, University of Potsdam, Potsdam, Germany}
\affil[3]{Institute of Physics and Astronomy, University of Potsdam, Potsdam, Germany}
\date{}

\begin{document}
\maketitle

\begin{abstract}
Ecological networks describe the interactions between different species, informing us of how they rely on one another for food, pollination and survival. If a species in an ecosystem is under threat of extinction, it can affect other species in the system and possibly result in their secondary extinction as well. Understanding how (primary) extinctions cause secondary extinctions on ecological networks has been considered previously using computational methods. However, these methods do not provide an explanation for the properties which make ecological networks robust, and can be computationally expensive. We develop a new analytic model for predicting secondary extinctions which requires no non-deterministic computational simulation. Our model can predict secondary extinctions when primary extinctions occur at random or due to some targeting based on the number of links per species or risk of extinction, and can be applied to an ecological network of any number of layers. Using our model, we consider how false positives and negatives in network data affect predictions for network robustness. We have also extended the model to predict scenarios in which secondary extinctions occur once species lose a certain percentage of interaction strength, and to model the loss of interactions as opposed to just species extinction. From our model, it is possible to derive new analytic results such as how ecological networks are most robust when secondary species degree variance is minimised. Additionally, we show that both specialisation and generalisation in distribution of interaction strength can be advantageous for network robustness, depending upon the extinction scenario being considered.

\end{abstract}

\section{Introduction}

No species exists in isolation, depending upon interactions with other species to feed, reproduce or maintain a stable population \cite{Haeckel1866,Cohen1978}. Modelling the interactions between species is therefore of great importance in ecology, and one approach to this problem is to model interactions as an ecological network \cite{Bersier2007,Ings2018}. Ecosystems are increasingly threatened by the effects of climate change \cite{Dawson2011,Bellard2012} which can cause sudden and widespread extinction events. Therefore, it is useful to model extinctions on ecological networks in order to understand the possible knock-on effects of species extinctions, as this may help to identify methods for conserving or reinforcing ecosystems in the future \cite{jongman1995,Forup2008,Tylianakis2010}.

Species extinctions on ecological networks have been extensively studied in the past 20 years, with simplistic topological models providing predictions for the impact of extinctions under scenarios including: extinctions which occur at random or with some ordering \cite{Memmott2004}, extinctions on networks made up of numerous trophic levels \cite{Pocock2012}, and extinctions which occur due to a loss of interaction strength over a certain threshold \cite{Schleuning2016}.

The models used are not the only possible approach to understanding the robustness of ecological networks. Other models consider the size of the largest component in the interaction network \cite{Sole2001,Montoya2006}, or take a more dynamical approach as is the case with Bayesian network models \cite{Aguilera2011,Ramazi2021}. Here we restrict ourselves to what we refer to as simplistic topological network models, which originate from Memmott et al. \cite{Memmott2004}. In these models, we are concerned with the point at which a given species goes extinct due to losing either a certain number of neighbours or a certain amount of interaction strength.

Previous work on simplistic topological network models have been largely computational, where extinctions are simulated in order to obtain predictions. Limited analytic work has been done to predict the robustness of ecological networks which are either maximally or minimally nested \cite{Burgos2007}, but there is no existing analytic framework which can predict the robustness of any given simple ecological network. In the following, we develop such a model, which improves upon computational methods by providing an insight into the properties that make ecological networks robust, and by cutting computational cost.

We start by considering the same scenario put forward by Memmott et al. \cite{Memmott2004}, where a bipartite mutualistic network (such as a plant pollinator network) undergoes extinctions on one trophic level, with species on the other trophic level experiencing secondary extinctions if they lose some or all of their neighbours. Secondary extinctions may be predicted for random or targeted primary extinctions, and secondary extinctions are predictable on networks with more than two trophic layers \cite{Pocock2012}. Our model may also be used to predict the effects of errors in network data, where interactions are erroneously included or excluded. The model is then developed further, taking into account the variable interaction strengths of neighbouring species, where a species will go extinct if it loses a certain amount of interaction strength, as considered by Schleuning et al. \cite{Schleuning2016}. We also consider the scenario in which species go extinct gradually, modelled by the loss of interaction strength as opposed to entire species.

Having developed an analytic model for these scenarios, we can determine the topological properties which make ecological networks robust. Previously, the roles of nestedness and specialisation have been debated as possible sources of robustness \cite{Memmott2004,Burgos2007,Nielsen2007}. We use our model to demonstrate that, when interaction strength is irrelevant, a network is most robust against random extinctions when the variance of its secondary species degree distribution is minimised. When interaction strength is included, we show that if secondary species' interaction strength is maximally specialised then network robustness is constant regardless of network degree distribution or extinction sensitivity. If interaction strength is maximally generalised, networks with high degree secondary species have robustness that is solely dependent upon extinction sensitivity. As a result, high specialisation makes a network more robust if it is highly sensitive to interaction loss, and high generalisation is better for robustness if interaction loss sensitivity is low.

\section{Introducing the Analytic Framework} \label{sec:rand_extinct}

\begin{figure}[h]
    \begin{minipage}{1\textwidth}
    \centering
    \captionsetup{justification=centering}
    \makebox[\textwidth]{\includegraphics[width = 1\textwidth]{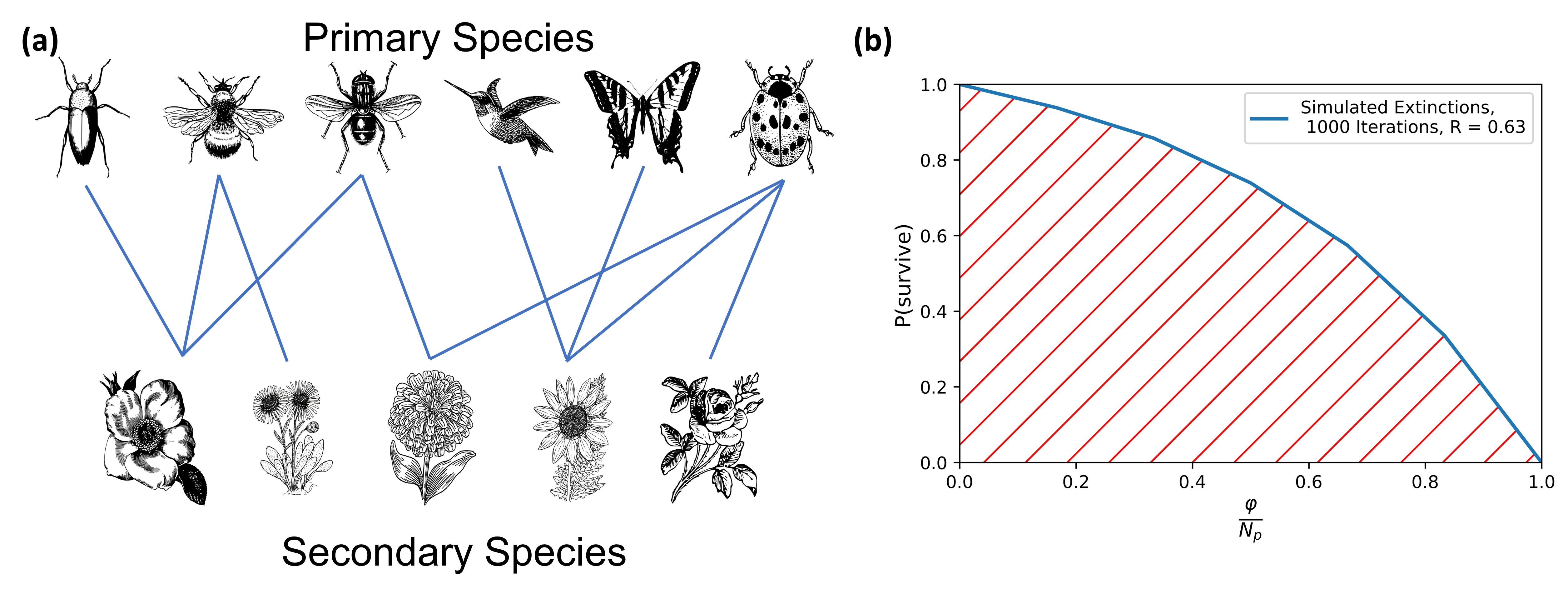}}
    \vspace{-20pt}
    \caption{\textbf{(a)} An example plant pollinator network and \textbf{(b)} its associated robustness curve. For this network, pollinators are treated as primary species, and plants as secondary species. Random primary extinctions are simulated repeatedly and the proportion of surviving secondary species is recorded in order to generate the robustness curve.}
    \label{fig:rob_example}
    \end{minipage}
\end{figure}

In the model of Memmott et al. \cite{Memmott2004}, species in one trophic level (e.g. pollinators) undergo extinctions, and this impacts species in an adjacent, secondary trophic level (e.g. plants). If species in the secondary level lose all of their neighbours, they suffer a secondary extinction. Primary extinctions may occur at random or according to some ordering, such as highest to lowest degree, where the degree of a species is the number of interactions/links/edges it has. We can plot the proportion of secondary species which survive against the proportion of primary extinctions in order to visualise how robust a given ecological network is against extinction, and an example of such a ``robustness curve'' is shown in Figure \ref{fig:rob_example}.

The area under the robustness curve in Figure \ref{fig:rob_example}\textbf{(b)} may be calculated in order to give a single metric for ecological network robustness, and this is given by \cite{Burgos2007}

\begin{equation} \label{eq:rob_eq}
    R = \frac{1}{N_p} \sum_{\varphi=0}^{N_p}  Pr(\text{survive}|\varphi),
\end{equation}

where $N_p$ is the total number of primary species, $\varphi$ is the number of primary species which have gone extinct at a given point, and $Pr(\text{survive}|\varphi)$ is the average probability of a randomly chosen secondary species surviving after some $\varphi$ primary species have been removed. Previously, calculations of the robustness curve and the robustness value $R$ have been done computationally, with some analytic results being derived for extreme cases \cite{Burgos2007}. As we show in the following, it is in fact possible to analytically predict the robustness curve of any given simple ecological network for a variety of extinction scenarios.

Let us consider some species $A$ in the secondary trophic level, which initially has degree $k_A$ and therefore $k_A$ unique neighbours in the primary level. If species in the primary level go extinct at random, we want to know the probability that species $A$ has degree $k_A - j$ (i.e. $j$ extinct neighbours) after some $\varphi$ number of primary species have gone extinct. If there are $N_p$ primary species, then there are ${N_p} \choose {\varphi}$ different possible combinations of primary species extinctions. We then need to find how many of those combinations include $j$ neighbours of $A$. There are $k_A \choose j$ possible combinations for removing $j$ neighbours of $A$, and therefore there are $N_p - k_A \choose \varphi - j$ possible combinations for removing $\varphi - j$ species which are not neighbours of $A$, so long as $\varphi \geq j$. Multiplying $k_A \choose j$ by $N_p - k_A \choose \varphi - j$ gives us the total number of combinations of length $\varphi$ which include $j$ neighbours of $A$, and so we may write the probability of $A$ having degree $k_A - j$ once $\varphi$ primary species are extinct as

\begin{equation} \label{eq:degree_k}
    Pr(k^{\prime}_{A} = k_A - j|\varphi) = 
    \begin{cases}
    \frac{{k_{A} \choose j}{N_{p} - k_{A} \choose \varphi - j}}{{N_{p} \choose \varphi}} &\text{if  } \varphi \geq j, \\
    0  &\text{otherwise},
    \end{cases}
\end{equation}

where $k^{\prime}_{A}$ refers to $A$'s actual degree value once $\varphi$ primary species have gone extinct. This is the hypergeometric distribution, which describes a process of sampling without replacement where each sample may pass (a neighbour of $A$ is removed) or fail (a non-neighbouring primary species is removed). If we specify that species $A$ goes extinct once its degree is $k^{\prime}_{A} = k_A - i_k$ or below (i.e. it has lost at least $i_k$ neighbours), then the disconnection probability for secondary species $A$ once some $\varphi$ primary species are extinct is

\begin{equation} \label{eq:dis_rand}
    Pr(A \text{ extinct}|\varphi) = \sum_{j = i_k}^{k_A} Pr(k^{\prime}_{A} = k_A - j|\varphi),
\end{equation}

Since extinction probability is only dependent upon the initial degree of a given secondary species species, the total number of primary species in the network and the number of primary species removed, we may extend this to all secondary species of initial degree $k$. Consequently, the average secondary extinction probability over the entire network is

\begin{equation} \label{eq:ave_dis}
    Pr(\text{extinct}|\varphi) = \sum_{k=0} p(k) \sum_{j = i_k}^{k} Pr(k^{\prime} = k - j|\varphi),
\end{equation}

where $p(k)$ is the probability of some randomly chosen secondary species having an initial degree of $k$. Given that $Pr(\text{survive}|\varphi)$ is simply $1-Pr(\text{extinct}|\varphi)$, we can rewrite the expression for robustness $R$ from Equation \ref{eq:rob_eq} as

\begin{equation}
    R = 1 - \frac{1}{N_p}\sum_{\varphi = 0}^{N_{p}} Pr(\text{extinct}|\varphi).
\end{equation}

Here we note that this analytic model is considerably computationally cheaper than brute force simulation. With an efficient implementation, calculating $Pr(\text{extinct}|\varphi)$ analytically takes $O(p)$ time, where $p$ is the number of unique non-zero entries in the secondary species degree distribution. By contrast, estimating $Pr(\text{extinct}|\varphi)$ computationally once takes $O(N_s)$ time, where $N_s$ is the number of secondary species and $N_s \geq p$. In practice, it is often necessary to run several thousand simulations in order to produce an accurate estimate of $Pr(\text{extinct}|\varphi)$, and so our analytic approach is substantially computationally cheaper than the brute force method.

\begin{figure}[h]
    \begin{minipage}{1\textwidth}
    \centering
    \captionsetup{justification=centering}
    \makebox[\textwidth]{\includegraphics[width = 1\textwidth]{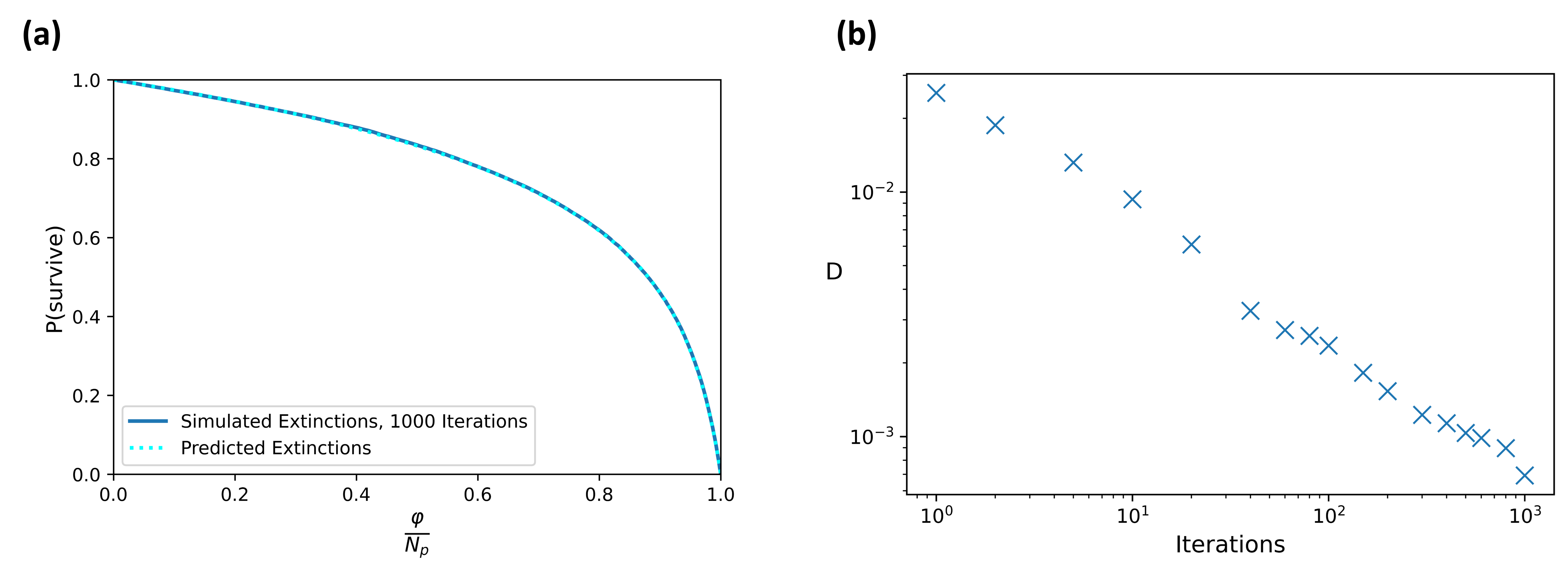}}
    \vspace{-20pt}
    \caption{\textbf{(a)} Analytically predicted and computationally simulated robustness curves of a real world network from a study by Kato \cite{kato2000} and \textbf{(b)} the absolute curve divergence between the analytic and simulated curve as computational simulations increase, plotted on a log-log scale.}
    \label{fig:eco_comparisons}
    \end{minipage}
\end{figure}

In Figure \ref{fig:eco_comparisons}\textbf{(a)}, we demonstrate the results of our model by comparing the analytically predicted robustness curve for an ecological network against the average curve obtained computationally when all neighbours must be removed for extinction to occur (i.e. $i_k = k$). The ecological data used is from a study of plant pollinator networks in Japan by Kato \cite{kato2000}. We can see that the computationally obtained curve converges to our predicted curve as the number of simulations increases, indicating that our method accurately predicts the average robustness curve. 

We also compare the absolute curve divergence between predicted and simulated curves for an increasing number of simulations in Figure \ref{fig:eco_comparisons}\textbf{(b)}. The absolute curve divergence $D$ is given by

\begin{equation}
    D = \frac{\sum_{\varphi = 0}^{N_{p}} |Pr(\text{survive}|\varphi)_{predict} - Pr(\text{survive}|\varphi)_{sim}|}{N_{p}},
\end{equation}

where $Pr(\text{survive}|\varphi)_{predict}$ and $Pr(\text{survive}|\varphi)_{sim}$ are the predicted and average simulated secondary species survival probabilities respectively. In Figure \ref{fig:eco_comparisons} we can see that the computationally generated result converges towards our prediction in the limit of a large number of simulations.

\section{Minimum Variance Maximises Robustness Against Random Extinctions}

Using this analytic model, we can prove that an ecological network is more robust when the secondary species degree distribution's variance is minimised, under the condition that secondary species average degree is held constant. In other words, the network is most robust if all secondary species have the same number of links to primary species as one another (or are as close to equal as possible). Let us consider the scenario in which a secondary species only goes extinct if it loses all of its primary neighbours. Now let us take two secondary species $A$ and $B$, where $k_A < k_B -1$, so if we were to remove an edge from $B$ and add it to $A$ they would be closer together in degree and $A$ would not have more edges than $B$. How would rewiring an edge like this affect their extinction probabilities, and by extension, the network's robustness? For a given number of extinct primary neighbours $\varphi$ we can write the change in average extinction probability for the network as

\begin{align}
    \Delta P &= Pr(\text{extinct}|\varphi)_{\text{rewired}} -  Pr(\text{extinct}|\varphi)_{\text{initial}} \nonumber \\ 
    &= Pr(k^{\prime}_{A} + 1 = 0|\varphi) - Pr(k^{\prime}_{A} = 0|\varphi) + Pr(k^{\prime}_{B} - 1 = 0|\varphi) - Pr(k^{\prime}_{B} = 0|\varphi) \nonumber \\ 
    &= \frac{1}{{N_{p} \choose \varphi}} \Bigg[ {N_{p} - (k_A + 1) \choose \varphi - (k_A + 1)} - {N_{p} - k_A \choose \varphi - k_A} + {N_{p} - (k_B - 1) \choose \varphi - (k_B - 1)} -{N_{p} - k_B \choose \varphi - k_B}\Bigg].
\end{align}

This may be simplified using the identity ${x \choose y} = {x-1 \choose y-1} + {x-1 \choose y}$ to give

\begin{equation}
    \Delta P = \frac{1}{{N_{p} \choose \varphi}} \Bigg[ {N_{p} - k_B \choose \varphi - (k_B - 1)} - {N_{p} - (k_A + 1) \choose \varphi - k_A} \Bigg],
\end{equation}

and if we expand the terms for ${N_{p} - k_B \choose \varphi - (k_B - 1)}$ and ${N_{p} - (k_A + 1) \choose \varphi - k_A}$ into their factorial forms, we can rearrange to give

\begin{equation}
    \Delta P = \frac{(N_{p} - k_B)!}{{N \choose \varphi} (\varphi - k_A)!(N-\varphi-1)!} \Bigg[\prod_{k=k_A+1}^{k_B-1} (\varphi - (k-1)) - \prod_{k=k_A+1}^{k_B-1} (N_{p} - k) \Bigg].
\end{equation}

Since $\prod_{k=k_A+1}^{k_B-1} (\varphi - (k-1)) \leq \prod_{k=k_A+1}^{k_B-1} (N_{p} - k)$ when $\varphi < N$ and $k_A < k_B -1$, we know that the change in disconnection probability $\Delta P \leq 0$ under the same conditions. Therefore, the disconnection probability must decrease or remain the same under the rewiring procedure, and so robustness must increase or remain the same. Additionally, we can prove that this rewiring will always reduce the variance of the secondary species degree distribution. The change in variance $\Delta \text{Var}$ is given by

\begin{align}
    \Delta \text{Var} &= E(k^2)_{\text{rewired}} - E(k^2)_{\text{initial}} \nonumber \\
    &= \frac{1}{N_s} \Big[ (k_A+1)^2 - k_A^2 + (k_B-1)^2 + k_B^2 \Big] \nonumber \\
    &= \frac{2}{N_s} (k_A - k_B + 1),
\end{align}

where $N_s$ is the number of secondary species in the network. From this, we know $\Delta \text{Var} < 0 $ when $k_A < k_B -1$, so variance always decreases for rewiring an edge from species $B$ to species $A$ given that initially $k_A < k_B -1$. This proves that, for a secondary species degree distribution of fixed degree, lower degree distribution variance entails higher robustness, and vice versa.

This result tells us exactly the structural properties that make secondary species in ecological networks robust against random primary species extinctions, namely equally distributed interactions. Previous research has indicated this before \cite{Burgos2007}, however, this was not conclusively proven, nor was the relationship between robustness and secondary species degree variance established. In the work of Burgos et al. \cite{Burgos2007}, robustness is related to nestedness, which is the propensity of primary species to interact with secondary species which other, higher degree primary species also interact with. Nestedness affects the degree distributions of both primary and secondary species, however, we know from our model that for robustness against random primary extinctions, the primary species degree distribution is irrelevant. Therefore, for random primary extinctions, nestedness is not necessarily an indicator of secondary species robustness. 

\begin{figure}[h]
    \begin{minipage}{1\textwidth}
    \centering
    \captionsetup{justification=centering}
    \makebox[\textwidth]{\includegraphics[width = 1\textwidth]{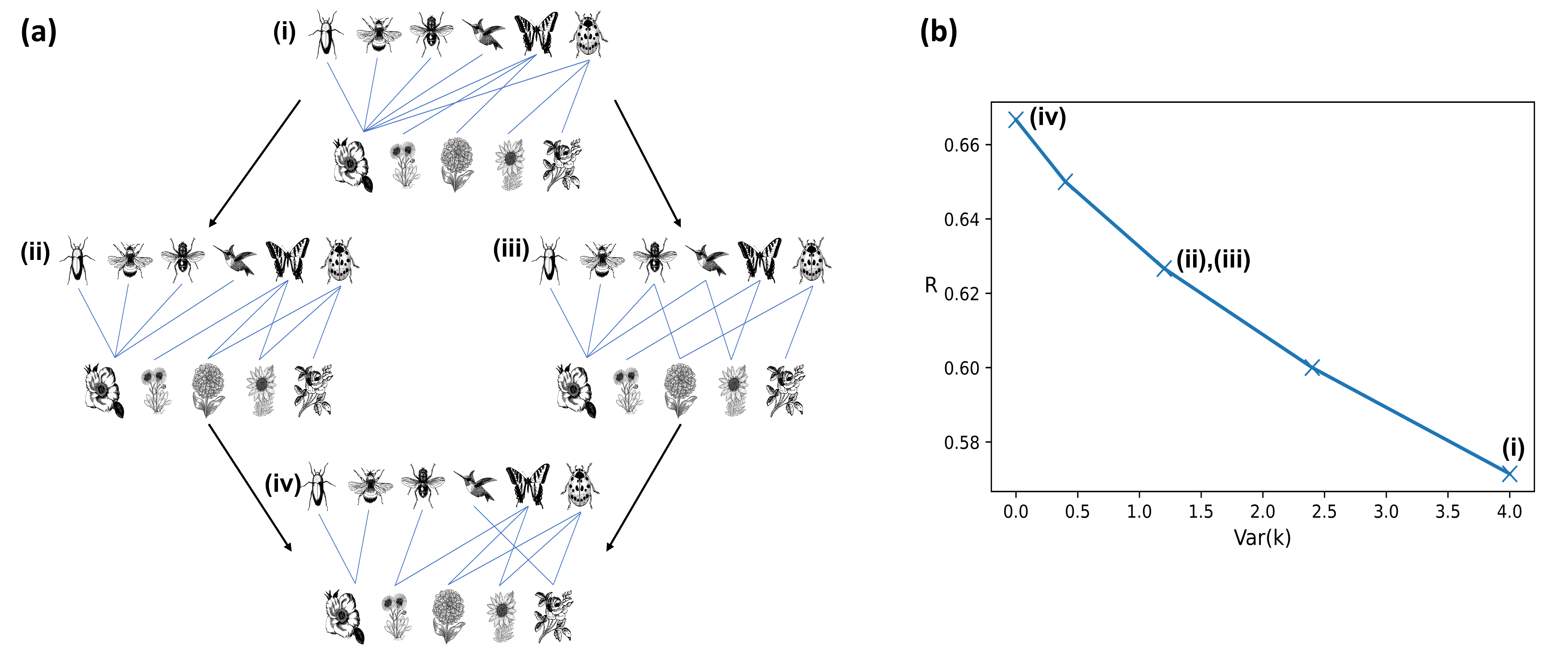}}
    \vspace{-20pt}
    \caption{\textbf{(a)} Example networks with different second neighbour degree variances and \textbf{(b)} network robustness plotted against second neighbour degree variance. Network (i) exhibits the highest variance and lowest robustness, networks (ii) and (iii) have the same variance and robustness even though they have different primary species degree distributions, and network (iv) has the lowest variance and highest robustness.}
    \label{fig:rob_var}
    \end{minipage}
\end{figure}

To illustrate the relation between degree variance and robustness, we provide a series of example networks in Figure \ref{fig:rob_var}, each with equal numbers of interactions and primary and secondary species, but different robustness and secondary degree distribution variance. We also show the correlation between robustness and secondary degree distribution variance for a network undergoing edge rewiring. The edge rewiring procedure starts on a network where a single secondary species is connected to all primary species, with all other secondary species having one primary neighbour, and one by one edges are swapped from the highest to lowest degree secondary species until secondary species degree variance is minimised.

We can see clearly that the highest variance network (i) has lowest robustness, and the lowest variance network (iv) has the highest robustness. Additionally, networks (ii) and (iii) have the same variance and robustness values as one another, as the only difference between them is the degree distribution of primary species. The difference in primary species degree distribution means that they are not considered to have the same nestedness as one another, but they are equally robust, demonstrating the fact that nestedness and robustness are not necessarily related.

\section{Targeted Species Extinctions}

\begin{figure}[h]
    \begin{minipage}{1\textwidth}
    \centering
    \captionsetup{justification=centering}
    \makebox[\textwidth]{\includegraphics[width = 0.4\textwidth]{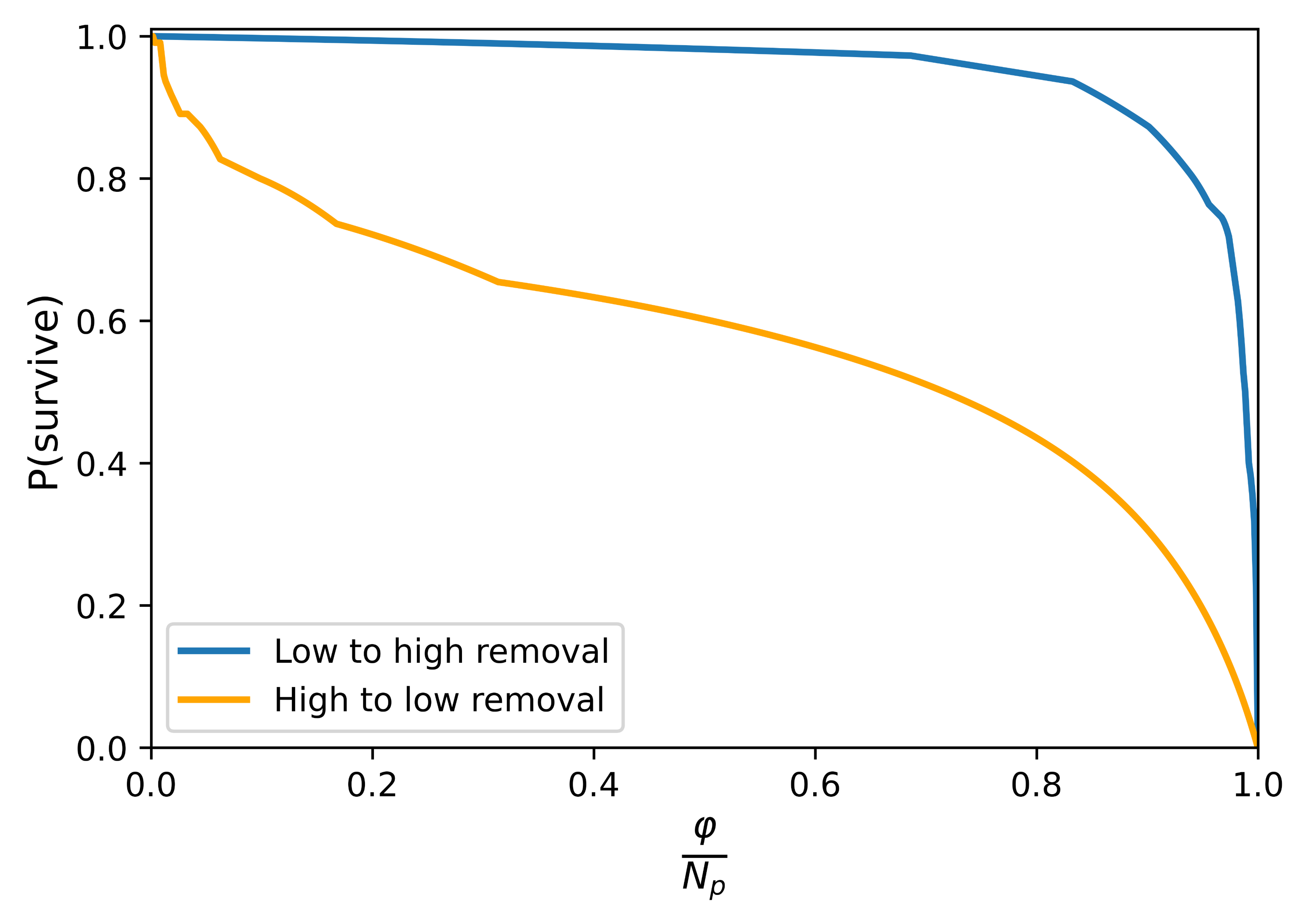}}
    \vspace{-20pt}
    \caption[Degree targeted species extinction on ecological networks]{Analytically predicted robustness curves for targeted primary species removal. Predictions are made on the plant pollinator network from a study by Kato \cite{kato2000}. The blue curve is for removal of primary species in ascending degree order (lowest degree first) and the orange curve is for removal of primary species in descending degree order (highest degree first).}
    \label{fig:high_low}
    \end{minipage}
\end{figure}

The model demonstrated in the preceding sections only predicts robustness when primary species are removed at random, but it is also possible to adjust the model to predict robustness when primary species are removed in descending or ascending degree order. This means that species with many links (descending order) or few links (ascending order) go extinct first. In these scenarios, primary species are effectively sorted into some $n$ different groups based on degree value. All species within a group are removed in a random order before moving onto the next group which is higher or lower in degree value, depending on the scenario. If we consider some secondary species, it will have some $k_l$ neighbours in a given primary species group where all primary species have degree $l$. As before, we set some threshold number $i_k$ of neighbouring species which must be lost before a given secondary species of degree $k$ goes extinct. We can then say that the secondary species will go extinct as we remove primary species from some group $d$ if it satisfies the conditions $\sum_{l=d}^{n} k_l \geq i_k$ and $\sum_{l=d+1}^{n} k_l < i_k$ for descending degree order removal, and $\sum_{l=0}^{d} k_l \geq i_k$ and $\sum_{l=0}^{d-1} k_l < i_k$ for ascending degree order removal. We can therefore write the probability of a some  secondary species $A$ going extinct when primary species are removed in descending degree order as

\begin{equation} \label{eq:high_degree}
    Pr(\text{A extinct}|\varphi) = 
    \begin{cases}
    0 &\text{if } \sum_{l=d}^{n} k_l < i_k \text{ or } \varphi_{d} < j_d,\\
 \frac{{k_{d} \choose j_d}{N_{d} - k_{d} \choose \varphi_d - j_d}}{{N_{d} \choose \varphi_d}} &\text{if  } \sum_{l=d}^{n} k_l \geq i_k \text{ and } \sum_{l=d+1}^{n} k_l < i_k \text{ and } \varphi_d \geq j_d, \\
    1  &\text{if } \sum_{l=d+1}^{n} k_l \geq i_k,
    \end{cases}
\end{equation}

where $N_d$ is the number of primary species in group $d$, $\varphi_d$ is the number of primary species removed from group $d$ and $j_d = i_k - \sum_{l=d+1}^{n} k_l$. Similarly, for primary species removal in ascending degree order we have

\begin{equation} \label{eq:low_degree}
    Pr(\text{A extinct}|\varphi) = 
    \begin{cases}
    0 &\text{if } \sum_{l=0}^{d} k_l < i_k \text{ or } \varphi_{d} < j_d,\\
 \frac{{k_{d} \choose j_d}{N_{d} - k_{d} \choose \varphi_d - j_d}}{{N_{d} \choose \varphi_d}} &\text{if  } \sum_{l=0}^{d} k_l \geq i_k \text{ and } \sum_{l=0}^{d-1} k_l < i_k \text{ and } \varphi_d \geq j_d, \\
    1  &\text{if } \sum_{l=0}^{d-1} k_l \geq i_k,
    \end{cases}
\end{equation}

where $j_k = i_k - \sum_{l=0}^{d-1} k_l$. The extinction probabilities $Pr(\text{A extinct}|\varphi)$ from Equations \ref{eq:high_degree} and \ref{eq:low_degree} can be averaged over the degree distribution of the network in a manner similar to Equation \ref{eq:ave_dis}, which allows us to predict the average extinction probability for any number of removed primary species. Therefore, we can predict the robustness curves for descending and ascending degree removal of primary species, and an example is given in Figure \ref{fig:high_low}.

\begin{figure}[h]
    \begin{minipage}{1\textwidth}
    \centering
    \captionsetup{justification=centering}
    \makebox[\textwidth]{\includegraphics[width = 1\textwidth]{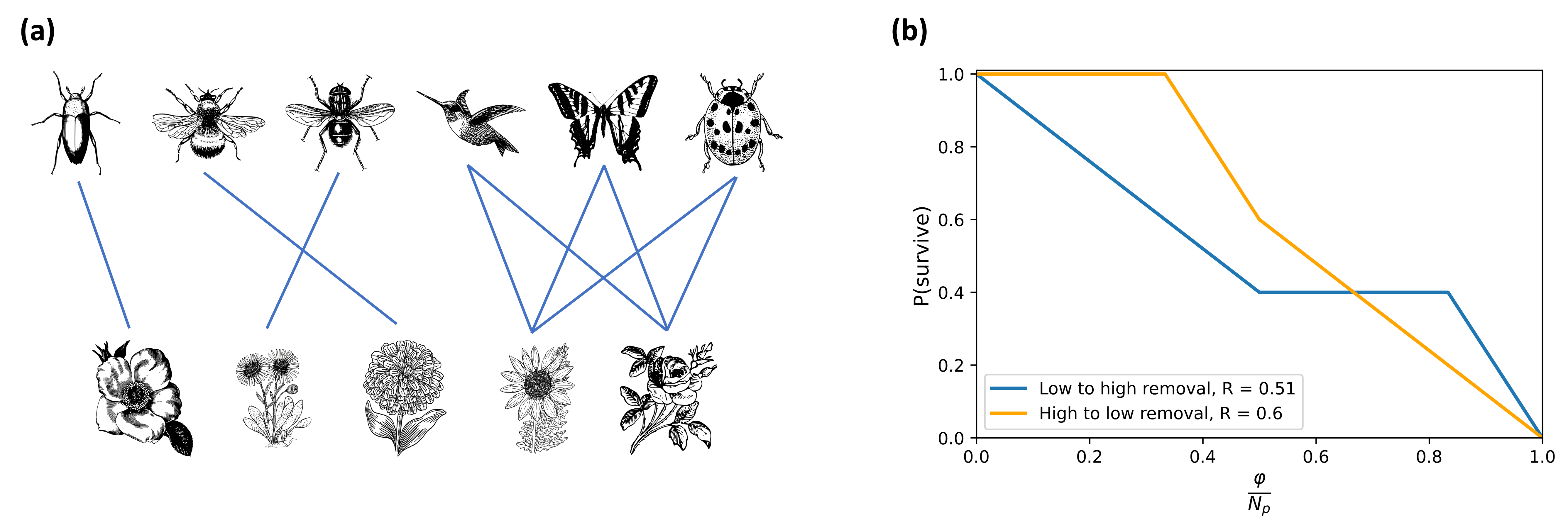}}
    \vspace{-20pt}
    \caption[Targeted species removal on an example ecological network]{\textbf{(a)} Example ecological network and \textbf{(b)} targeted removal robustness curves. Removing low degree primary species (pollinators) first gives a lower robustness than removing high degree primary species.}
    \label{fig:eco_scheme}
    \end{minipage}
\end{figure}

As before, our analytic model can successfully predict the average extinction probabilities for secondary species as primary species are removed. The scenarios in which primary species are removed in descending and ascending degree order have been referred to as the ``worst'' and ``best'' case scenarios respectively. However, our analytic model suggests that from the perspective of robustness, this may not exactly be the case. Under descending degree order removal, a secondary species' extinction probability only depends upon how many lowest degree neighbours it has, and for ascending degree order removal it depends upon the number of highest degree neighbours. In Figure \ref{fig:eco_scheme}, we provide an example ecosystem for which descending degree order removal gives higher network robustness than ascending degree removal.

While this is a specifically constructed example, it demonstrates that finding the true worst or best case scenario for secondary extinctions is not necessarily a case of removing primary species in descending or ascending degree order respectively. As such, a possible future line of enquiry is to try and establish the true worst and best case scenarios for species extinction on any given network.

Thusfar in this section we have considered species extinctions which are targeted based on degree value, but this is only one possible extinction ordering. Recent research has examined extinction scenarios in which species are lost according to their extinction risk as assessed by the IUCN (International Union for Conservation of Nature) Red List \cite{lamperty2022loss}. Species are ranked in the Red List from Critically Endangered to Least Concern, and in work by Lamperty and Brosi \cite{lamperty2022loss} frugivore species in a seed dispersal network are removed from highest to lowest extinction risk. Since species only belong to one of these extinction risk categories, it is necessary to simulate extinctions from each risk category in descending risk order, with the order of extinctions within each group randomised. This framework fits well with our targeted species extinction model.

\begin{figure}[h]
    \begin{minipage}{1\textwidth}
    \centering
    \captionsetup{justification=centering}
    \makebox[\textwidth]{\includegraphics[width = 0.4\textwidth]{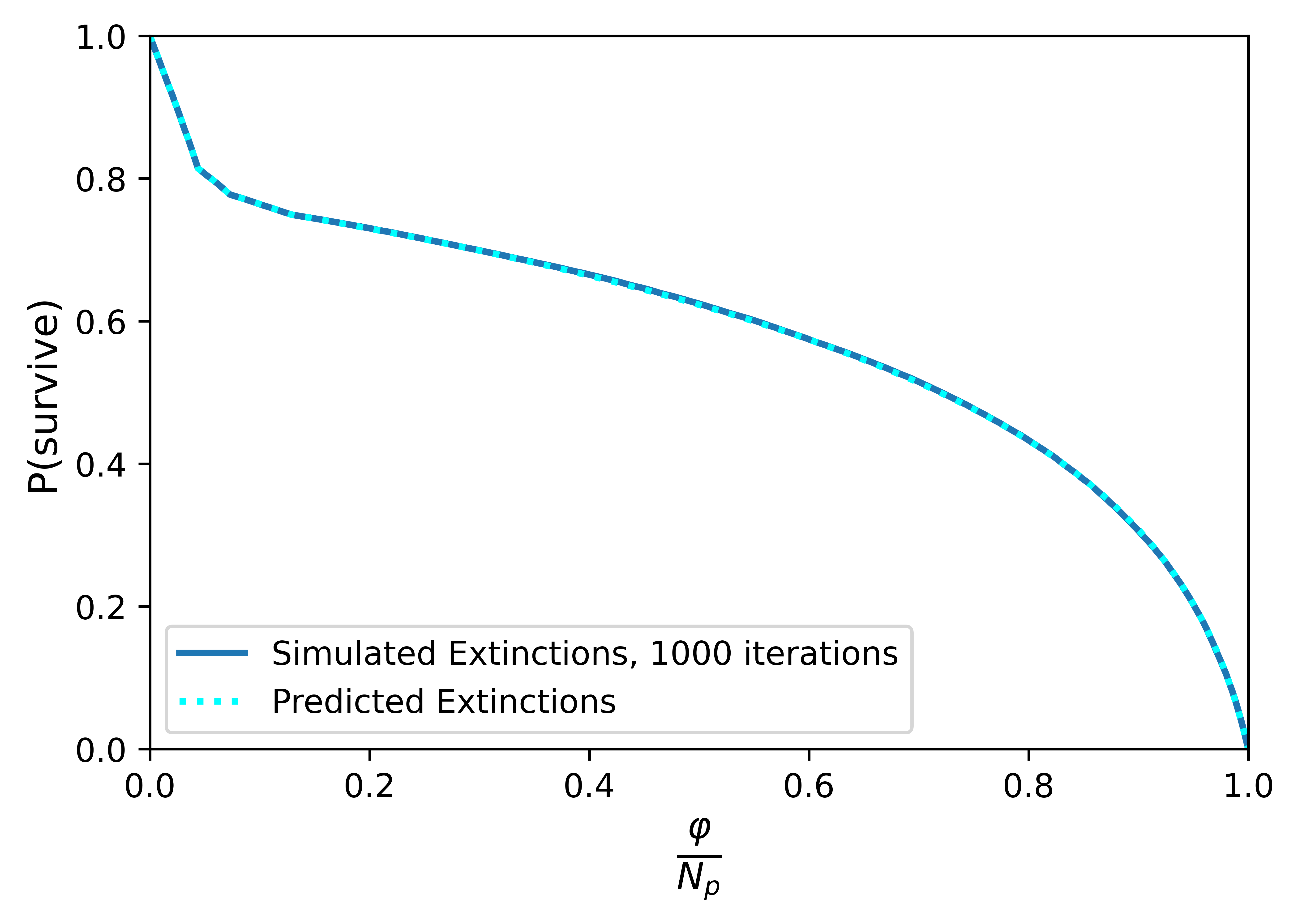}}
    \vspace{-20pt}
    \caption{Analytically predicted and computationally simulated robustness curves of a real world seed dispersal networks, where primary frugivore species go extinct according to their IUCN extinction risk. This replicates results from Lamperty and Brosi \cite{lamperty2022loss} using data from Bello et al. \cite{Bello2017}.}
    \label{fig:iucn_extinct}
    \end{minipage}
\end{figure}

Using data from Bello et al. \cite{Bello2017}, we can replicate the results of Lamperty and Brosi \cite{lamperty2022loss}, predicting the survival of plant species in a seed dispersal network as frugivore species are lost in descending extinction risk order. We calculate extinction probability in the same way as Equation \ref{eq:high_degree}, except instead of our primary species groups being organised by degree value, they are now organised by extinction risk. In Figure \ref{fig:iucn_extinct}, we show simulated and predicted plant species survival probabilities as frugivore species are removed in descending extinction risk order, with plant species going extinct once they have lost all of their frugivore neighbours.

This demonstrates the fact that our framework for targeted species extinctions may be extended to any ordering of primary species loss where primary species are sorted into groups which go extinct in some order, but extinction within groups occurs at random.

\section{Multi-layer Ecosystems}

So far, we have only demonstrated our model for bipartite systems, i.e. those which include only two groups that interact with one another. However, real world ecosystems can exist on several distinct layers, for example, predators may feed on pollinators which in turn pollinate plants. Another extension for our model is to predict species extinction in a group of species not directly adjacent to the group undergoing primary extinction. This is predictable analytically, but only for the scenario in which a species must lose all of its neighbours in order to go extinct.

Scenarios such as this have previously been considered by Pocock et al. \cite{Pocock2012}, using computational methods. Here we construct an example network to demonstrate robustness predictions on multi-layer networks. Let us consider a system of plants, pollinators and predators, where predators feed on pollinators, who in turn feed on plants. We want to know the probability of a predator going extinct after a certain number of plant extinctions. For some predator species A, the species will go extinct if all of the pollinator species it is connected to go extinct, which only occurs once all of their plant species neighbours go extinct. Therefore, the extinction probability of predator species A is simply dependent upon the $u_A$ initial number of unique plant species to which it is connected via its pollinator species neighbours. Therefore, we can treat the predator species in this system as our secondary species and the plant species as primary species.

Similar to Equation \ref{eq:degree_k}, the extinction probability for a secondary species A once some $\varphi$ number of primary species have been removed is

\begin{equation} \label{eq:dis_layer}
    Pr(u^{\prime}_{A} = 0|\varphi) = 
    \begin{cases}
    \frac{{N_{p} - u_{A} \choose \varphi - u_{A}}}{{N_{p} \choose \varphi}} &\text{if  } \varphi \geq u_A, \\
    0  &\text{otherwise},
    \end{cases}
\end{equation}

\begin{figure}[h]
    \begin{minipage}{1\textwidth}
    \centering
    \captionsetup{justification=centering}
    \makebox[\textwidth]{\includegraphics[width = 1\textwidth]{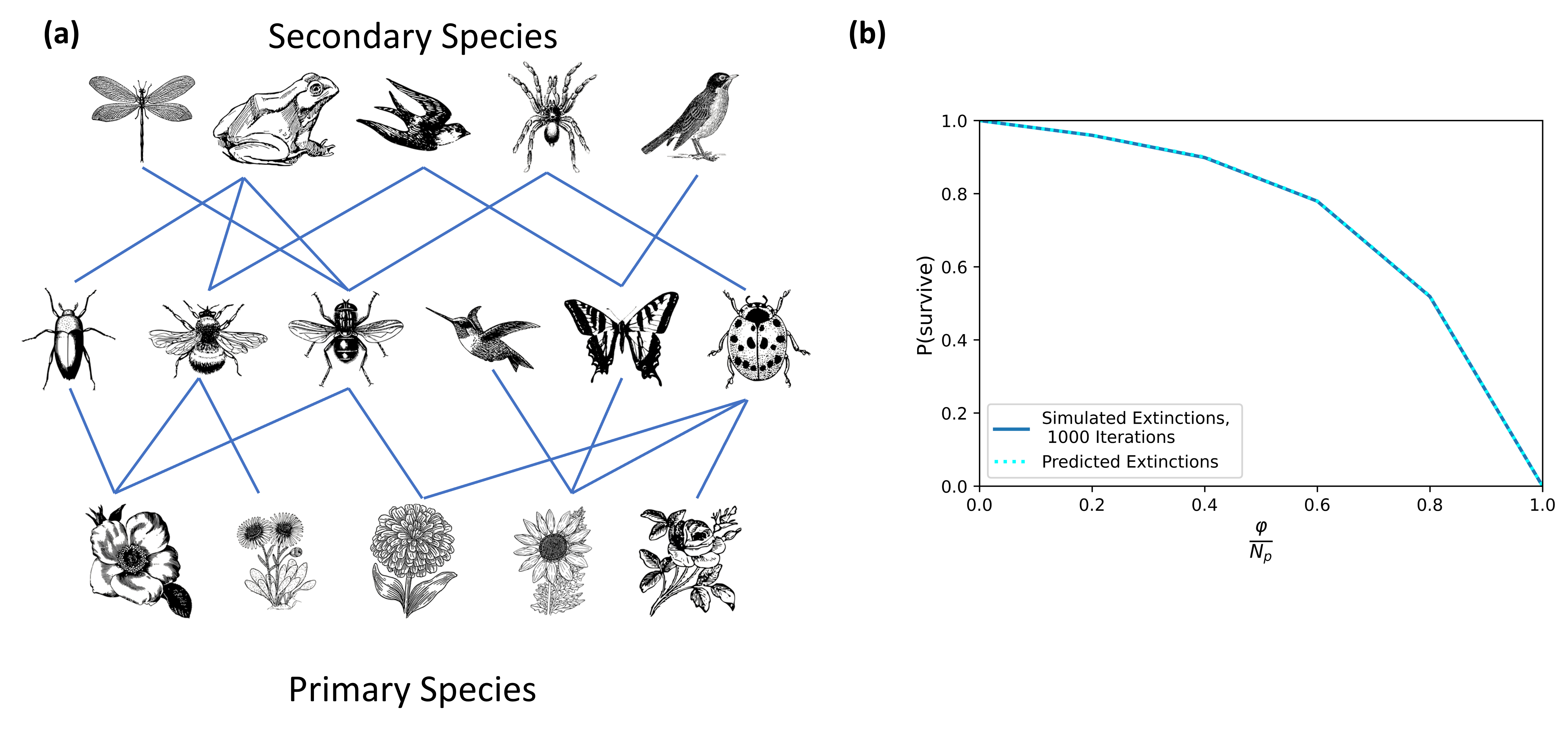}}
    \vspace{-20pt}
    \caption{\textbf{(a)} Example three layer network of plants (primary species), pollinators and predators (secondary species), and \textbf{(b)} the associated analytically predicted and computationally simulated robustness curves.}
    \label{fig:layer_predict}
    \end{minipage}
\end{figure}

As before, we can average Equation \ref{eq:dis_layer} over the distribution of secondary species connected to $u$ unique primary species to predict secondary extinctions as primary species are removed. The analytically predicted and computationally simulated robustness curves for this are given in Figure \ref{fig:layer_predict}.

While this expands the scope of our analytic model beyond simply two layer ecosystems, it is important to note that thusfar we can only model extinctions on multi-layer systems if species go extinct after losing all of their neighbours. Therefore, we cannot consider as many different extinction scenarios on multi-layer networks than we can for bipartite networks.

\section{False Positives and Negatives in Network Data}

Beyond simply predicting robustness, we may also be interested in how predictions of robustness are affected by errors in network data, as ecological data can be error prone \cite{Kangas2018,Aguiar2019}. For example, networks may vary across environmental gradients \cite{pellissier2018comparing} or may constitute metawebs inferred from proxies \cite{morales2015inferring,maiorano2020tetra}. One may be interested in how the robustness of networks change as false edges are added in (false positives) or true edges are removed (false negatives). In the simplest case, let us consider the random addition and removal of edges. Since robustness against random primary species removal only depends upon the degree distribution of secondary species, we can analytically predict how robustness will change as edges are randomly added or removed by modelling the changes to the secondary degree distribution.

For random edge addition and removal, we can define recursive formulae which describe how the secondary degree distribution will change. For random edge addition, the recursive formula which describes the probability of randomly choosing a secondary species with degree $k$ after some $t$ edges have been added is

\begin{align}
    p(k)_t = p(k)_{t-1} &- p(k)_{t-1} \frac{N_{p} - k}{N_{p}N_{s} - (E + t-1)} \\ \nonumber 
    & + p(k-1)_{t-1} \frac{N_{p} - (k-1)}{N_{p}N_{s} - (E + t-1)},
\end{align}

where $E$ is the total number of edges in the network before any additional edges have been added. Note that $N_{p}N_{s} - (E + t-1)$ is the total possible number of edges which could be added to the network once $t-1$ edges have been added.

For random edge removal, the recursive formula for the probability of choosing a secondary species with degree $k$ after some $v$ edges have been removed is

\begin{equation}
    p(k)_v = p(k)_{v-1} - p(k)_{v-1}\frac{k}{E - (v-1)} + p(k+1)_{v-1}\frac{k+1}{E - (v-1)}.
\end{equation}

In order to simulate the presence of false positives or negatives in network data, we can apply these recursive formulae a certain number of times in order to adjust the degree distribution, and then assess the impact of false positives and negatives by predicting robustness for either scenario.

\begin{figure}[h]
    \begin{minipage}{1\textwidth}
    \centering
    \captionsetup{justification=centering}
    \makebox[\textwidth]{\includegraphics[width = 1.2\textwidth]{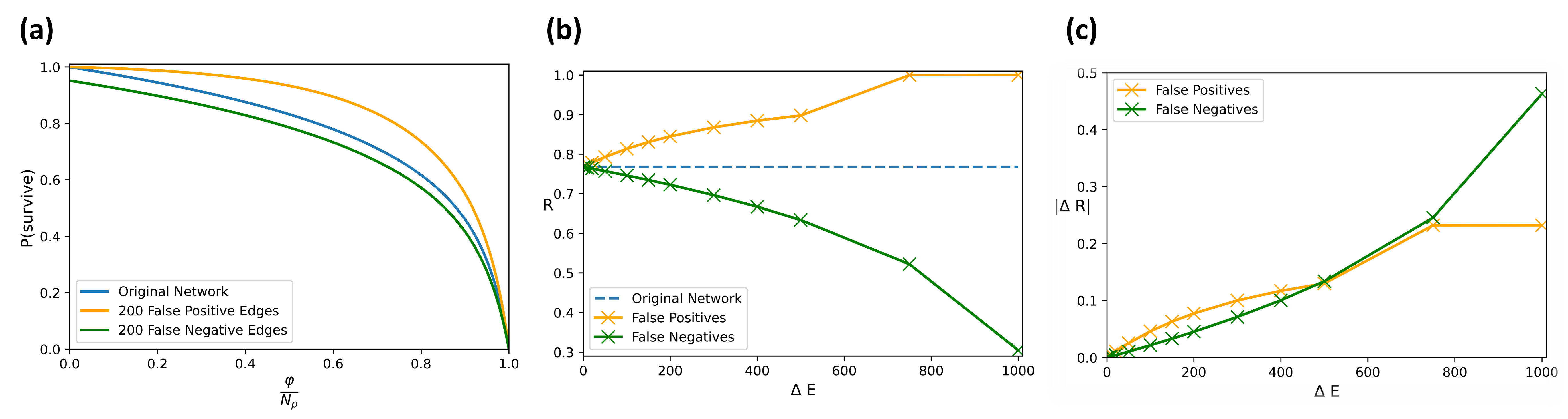}}
    \vspace{-20pt}
    \caption{\textbf{(a)} Analytically predicted robustness curves for a real world ecological network from a study by Kato \cite{kato2000} for the original network, the network with 200 edges randomly added and the network with 200 edges randomly removed. The original network has 1125 edges. \textbf{(b)} Shows how network robustness changes as edges are added or removed, and \textbf{(c)} shows the difference between the robustness of networks with errors and the original networks as edges are added or removed.}
    \label{fig:eco_false}
    \end{minipage}
\end{figure}

In Figure \ref{fig:eco_false}, we show the robustness curve for a plant pollinator network undergoing random extinctions where $i_k = k$, comparing the original predicted curve against the predicted curves for including false positives and negatives.


In terms of robustness, false positives have a more significant impact than false negatives in small quantities. The network we use to generate the data shown in Figure \ref{fig:eco_false} has 1125 edges, and we see that up to a change in edges $\Delta E = 500$ (roughly $44\%$ the original number of edges) false positives increase robustness more than false negatives decrease it. For false positives, as $R \rightarrow 1$, additional edges contribute less and less to robustness and for false negatives, as $\Delta E \rightarrow E$, $R \rightarrow 0$, so we see a larger change in robustness due to false negatives for large numbers of errors. We verify similar results on a dataset of several networks, and details of this data from 18 real world plant pollinator networks is given in the Supplementary Materials. For this dataset, we find that if we measure robustness where $20\%$ of the original number of edges are added/removed for false positives/negatives, then the net change in robustness is always positive, i.e. false positives always increase robustness more than false negatives decrease it.

If we assume that ecological network data gathering in the real world is reasonably accurate, i.e. unlikely to over/under record interactions by more than $20\%$, then we would expect false positives to introduce more error into calculations of robustness than false negatives. Particular care is needed for robustness analyses based on metawebs of potential trophic interactions for which the false positive and false negative rates are difficult to ascertain \cite{morales2015inferring,maiorano2020tetra}.

While this result indicates that false positives have more impact than false negatives, it only provides one perspective for how these errors may be introduced into network data. One future avenue of enquiry is to establish the likely sources of errors and model those, as opposed to modelling errors randomly.


\section{Species Specialisation and Generalisation}

In previous sections, we have only considered networks in which interactions are weighted equally, i.e. each one of a secondary species' interactions is as important for its survival. However, on real ecological networks, a secondary species may interact more with one primary species than another, and this has an impact on a secondary species survivability \cite{Berlow1999}. We can specify a certain percentage of total interaction strength that a species must lose before it goes extinct, an approach used before before by Schleuning et al. \cite{Schleuning2016}. We can update our extinction probability for some secondary species A to

\begin{equation}
    Pr(A \text{ extinct}|\varphi) = \sum_{j = 1}^{k_A} Pr(k^{\prime}_{A} = k_A - j|\varphi) Pr(\sum_{0}^j W \geq i_A),
\end{equation}

where $W$ is a random variable representing some randomly chosen weight corresponding to the interaction strength with a neighbour of $A$, $i_A$ is the interaction strength threshold for $A$ that must be removed before $A$ goes extinct, and $Pr(\sum_0^j W \geq i_A)$ is the probability of choosing some $j$ weighted interactions which exceed the threshold $i_A$. We can get the value of $i_A$ from specifying some ratio of interaction strength that must be lost for secondary extinction to occur as the sensitivity threshold $T$, and calculating $i_A = \ceil{T k_A}$. The notation of $\ceil x$ refers to $x$ being rounded up to the nearest integer. The extinction probability $Pr(A \text{ extinct}|\varphi)$ may be averaged over all species then over values of $\varphi$ to obtain a robustness value for the network.

There is no closed form solution for $Pr(\sum_0^j W \geq i_A)$ since the weights $w_z$ do not necessarily follow a particular distribution. It is instead necessary to estimate $Pr(\sum_0^j W \geq i_A)$ in some way. The brute force method is to randomly sample $j$ weights using a Monte Carlo method, however, this must be repeated many times in order to give an accurate estimate, and is subject to statistical fluctuations. 

Instead, we have developed a deterministic sampling method, where a species' weights $w_z$ are arranged in size order and assigned a variable $y_z$, which takes values of 0 or 1. For $j$ removals there will be $j$ values of $y_z = 1$, with the rest equal to 0. We can express the sequence of weights as a sequence of 0 or 1 $y_z$ values, giving us a binary number. If weights were ordered as powers of 2, i.e. $w_z = 2^{k_A-z}$, then we could find the $n^{th}$ binary sequence of $y_z$ values with $j$ values equal to 1 above which all $\sum_{z=0}^{k_A} w_z y_z \geq i_A$ and below which all $\sum_{z=0}^{k_A} w_z y_z < i_A$, allowing us to calculate $Pr(\sum_0^j W \geq i_A)$ exactly. However, weights are not typically ordered as powers of two, so finding an exact result this way is rarely possible. Instead we can order binary sequences of $y_z$ values and sample from these orderings at some specified ``depth" in order to estimate $Pr(\sum_0^j W \geq i_A)$, where depth effectively determines how many samples are taken. This sampling method is fully deterministic, so for a given sequence of weights and a specified depth, we always return the same estimate for $Pr(\sum_0^j W \geq i_A)$. Further details about how this algorithm operates are given in the Supplementary Materials.

\begin{figure}[h]
    \begin{minipage}{1\textwidth}
    \centering
    \captionsetup{justification=centering}
    \makebox[\textwidth]{\includegraphics[width = 1.2\textwidth]{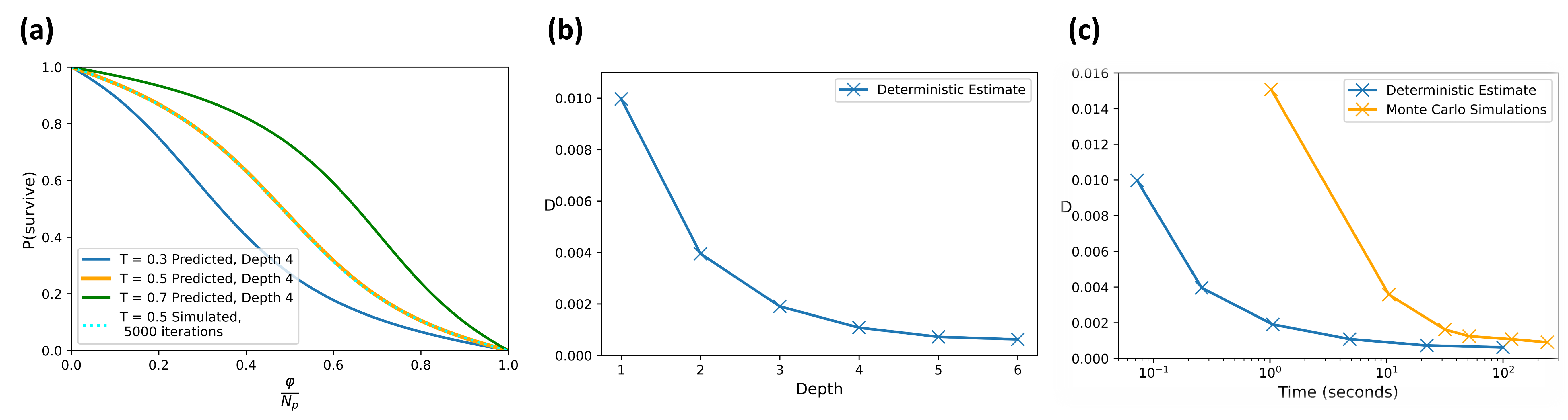}}
    \vspace{-20pt}
    \caption{\textbf{(a)} Analytically predicted and computationally simulated robustness curves for a real world network from a study by Kato \cite{kato2000} where unevenly weighted interaction strength is taken into account and extinctions occur over a specified threshold $T$ of interaction strength loss. Analytic predictions are given for threshold values of $70\%,50\%$ and $30\%$, and computationally simulated curve is given for $50\%$. \textbf{(b)} Comparison between depth of the estimation for $Pr(\sum_0^j W \geq i_A)$ and divergence between analytically predicted robustness curve and the simulated curve averaged over 5000 iterations. \textbf{(c)} Divergence between the predicted robustness curve and the 5000 iteration simulation curve compared against the time taken, with data for both the deterministic estimation method and the Monte Carlo method.}
    \label{fig:inter_strength}
    \end{minipage}
\end{figure}

Using our deterministic sampling method, we can provide quasi-analytic predictions for secondary species survival on networks undergoing random primary extinctions, where interaction strength is weighted unevenly and secondary extinctions occur after the loss of a certain percentage of interaction strength. Example predictions are given in Figure \ref{fig:inter_strength}, alongside results showing how our deterministic estimate becomes increasingly accurate with greater depth, and a comparison between the time taken to estimate $Pr(\sum_0^j W \geq i_A)$ and prediction accuracy for our deterministic sampling method and for a brute force Monte Carlo method.

From this, we can see that as the depth of the deterministic estimation increases, we get diminishing returns in terms of prediction accuracy, and that it is more computationally efficient to use the deterministic estimation method as opposed to Monte Carlo simulation in order to get the same level of prediction accuracy.

Having developed an analytic framework for secondary species extinctions when interaction strength is weighted unevenly, we can examine some extreme scenarios of interaction strength weighting. One property of interest in ecological networks is specialisation \cite{Bluthgen2006}, where specialist species tend to interact with a small number of species very strongly, and generalist species tend to interact with many species evenly. Given an ecological network with a set number of primary and secondary species, and a set distribution of interactions, we can examine the most specialist interaction weighting and the most generalist interaction weighting.

In the most specialist case, each secondary species weights one of its interactions at close to $100\%$ of its interaction strength, and all others close to $0\%$. Therefore, a given secondary species $A$ only goes extinct when it loses the neighbour with which it shares almost all interaction strength. If a neighbour of $A$ goes extinct, the probability of losing the heavily weighted neighbour is simply $\frac{1}{k_A}$, so $Pr(\sum_0^j W \geq i_A) = \frac{j}{k_A}$. This give an extinction probability for some species $A$ of

\begin{align}
    Pr(A \text{ extinct}|\varphi) & = \sum_{j = 1}^{k_A} Pr(k^{\prime}_{A} = k_A - j|\varphi) \frac{j}{k_A}, \nonumber \\
    & = \frac{\mathbb{E}(k^{\prime}_A)}{k_A} \nonumber \\
    & = \frac{\varphi}{N_p},
\end{align}

which we derive from the fact that $\mathbb{E}(k^{\prime}_A) = k_A \frac{\varphi}{N_p}$ since $Pr(k^{\prime}_{A} = k_A - j|\varphi)$ describes the hypergeometric distribution. This results in a robustness value of $R = 0.5$, regardless of the secondary degree distribution, number of primary species or threshold.

For the most generalist case, each secondary species weights all of its interactions evenly, which means $Pr(\sum_0^j W \geq i_A) = 0$ when $j < i_k$, and $Pr(\sum_0^j W \geq i_A) = 1$ when $j \geq i_k$. Therefore, $Pr(A \text{ extinct}|\varphi) = \sum_{j = i_k}^{k_A} Pr(k^{\prime}_{A} = k_A - j|\varphi)$, the same as Equation \ref{eq:dis_rand}.

Given these results, when is it more advantageous for a network to be highly specialist or highly generalist in terms of robustness? Let us consider some secondary species $A$ which is connected to all primary species in its network, i.e. $k_A = N_p$. Therefore, the extinction probability for $A$ is given by

\begin{align}
    Pr(A \text{ extinct}|\varphi) & = \sum_{j = i_k}^{k_A} 
    \begin{cases}
    \frac{{N_{p} \choose j}{0 \choose \varphi - j}}{{N_{p} \choose \varphi}} &\text{if  } \varphi \geq j, \\
    0  &\text{otherwise},
    \end{cases} \nonumber \\
    & = \begin{cases}
    1 &\text{if  } \varphi \geq i_k, \\
    0  &\text{otherwise},
    \end{cases}
\end{align}

\begin{figure}[h]
    \begin{minipage}{1\textwidth}
    \centering
    \captionsetup{justification=centering}
    \makebox[\textwidth]{\includegraphics[width = 1\textwidth]{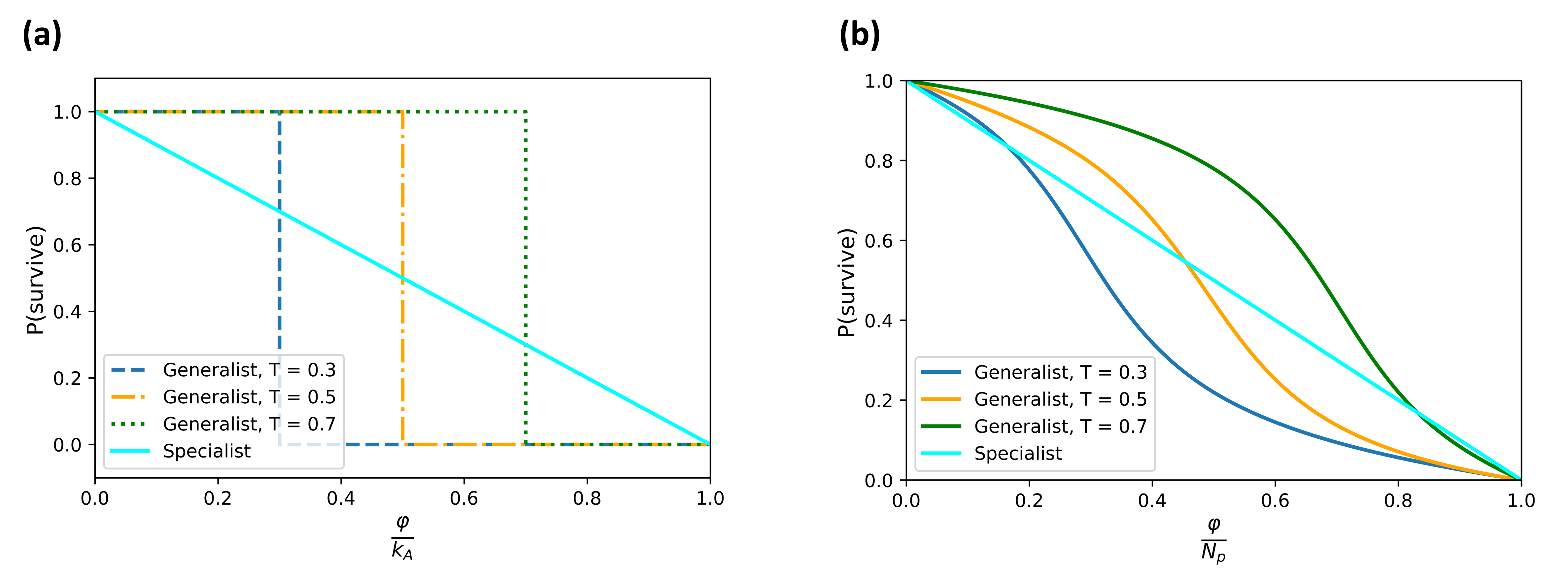}}
    \vspace{-20pt}
    \caption{\textbf{(a)} Shows species survival for maximal generalists and maximal specialists at various sensitivity thresholds $T$ when $k_A = N_p$. \textbf{(b)} Gives robustness curves for maximal generalisation and maximal specialisation on a real world network from a study by Kato \cite{kato2000} at various sensitivity thresholds $T$.}
    \label{fig:gen_spec}
    \end{minipage}
\end{figure}

If all secondary species in a network have $k = N_p$, then when they are maximally generalist, the network robustness is $R = \frac{i_k}{N_p}$. Therefore, such a network is more robust when secondary species are maximally generalist if more than $50\%$ of interaction strength must be lost before a secondary species goes extinct, i.e. when the sensitivity threshold $T > 0.5$. Conversely, the network is more robust when secondary species are maximally specialist if less than $50\%$ of interaction strength must be lost to make secondary species go extinct, so $T < 0.5$. To illustrate this, we provide robustness curves in Figure \ref{fig:gen_spec} of \textbf{(a)} a single secondary species with $k_A = N_p$, and of \textbf{(b)} an entire real world network.

From these results, we know that either extreme of species specialisation can be advantageous from the perspective of maximising network robustness, depending upon the sensitivity of the network, i.e. the proportion of interaction strength that must be lost for secondary species to go extinct. However, we see in Figure \ref{fig:gen_spec} \textbf{(b)} that it is not strictly the case on real networks that maximum generalisation is always better for robustness than maximum specialisation when $T > 0.5$, as the maximum generalist curve when $T = 0.5$ gives $R = 0.471$. This is due to the fact that secondary species typically have $k < N_p$ on real networks. Additionally, we note that the robustness values from the maximum generalist and maximum specialist interaction weightings do not necessarily give the maximum and minimum robustness values for a given threshold. Nevertheless, these results are still indicative of the fact that species generalisation and specialisation can both improve network robustness in different contexts, and so we might expect that in the real world, a network that has developed to be highly generalist is less sensitive to interaction loss than a network which has developed in order to be highly specialist.

\section{Interaction Loss}

The models we have considered are only concerned with the loss of primary species as a whole, where a primary species is removed at each ``step'' in the extinction process. Considering the loss of entire species at a time can skew our understanding of network robustness, for example a plant animal network with more animal than plants will appear more robust against primary extinctions of animal than against primary extinctions of plants \cite{Schleuning2016}. However, is this a realistic understanding of how species go extinct? There may be more animal species than plant species, but what if there is a very large population of each plant species and a small population of each animal species? Extinctions may be experienced more gradually, where a species' population dies off over time rather than all at once \cite{Valiente-Banuet2015}. This process can be modelled by examining the loss of interactions as opposed to the loss of species, which in network terms entails considering edge removal as opposed to node removal.

\begin{figure}[h]
    \begin{minipage}{1\textwidth}
    \centering
    \captionsetup{justification=centering}
    \makebox[\textwidth]{\includegraphics[width = 0.5\textwidth]{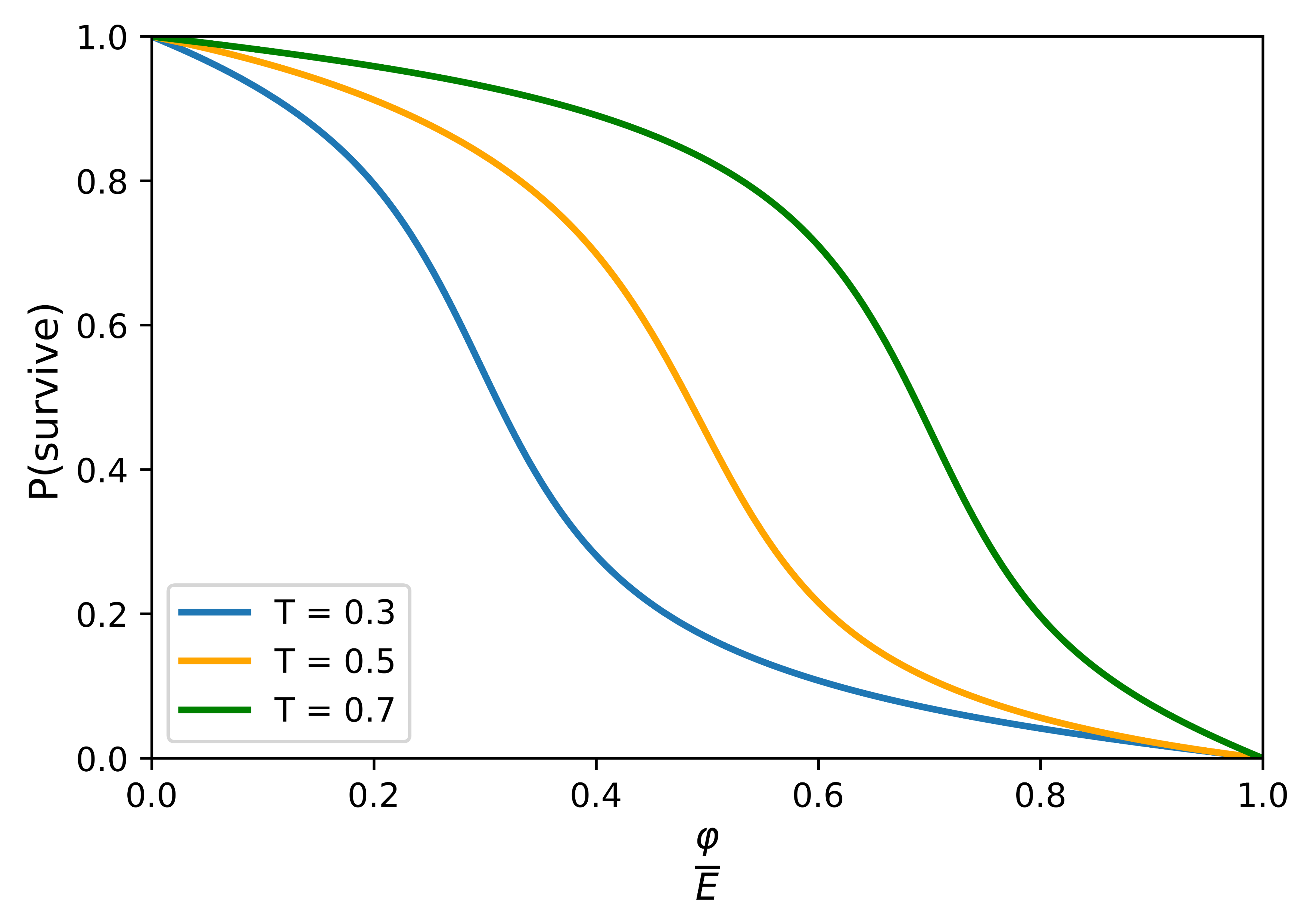}}
    \vspace{-20pt}
    \caption{Robustness curves for interaction loss on a real world network from a study by Kato \cite{kato2000} at varying sensitivity thresholds $T$.}
    \label{fig:bond_remove}
    \end{minipage}
\end{figure}

If interaction strength between species is represented as integer values, then we can treat each unit of interaction strength as an edge, so secondary species have degree values equal to the sum of their interaction strength with other species. We then have $E$ ``edges'' (i.e. total interaction strength on the network), and we remove some $\varphi$ units of interaction strength. For a given secondary species $A$, after removing some $\varphi$ interaction strength the probability that it has lost some interaction strength $j$ is

\begin{equation}
    Pr(k^{\prime}_{A} = k_A - j|\varphi) = 
    \begin{cases}
    \frac{{k_{A} \choose j}{E - k_{A} \choose \varphi - j}}{{E \choose \varphi}} &\text{if  } \varphi \geq j, \\
    0  &\text{otherwise},
    \end{cases}
\end{equation}

where, as before in the case of Equation \ref{eq:degree_k}, $k_{A}$ is the initial degree/total interaction strength of $A$, and $k_{A}^{\prime}$ is the degree/total interaction strength of $A$ after removing $\varphi$ interaction strength. From this, it is straightforward to predict secondary species' survival probability and network robustness using a similar logic as in Section \ref{sec:rand_extinct}. Predictions for interaction loss on an ecological network are given in Figure \ref{fig:bond_remove}.

These predictions of interaction loss give different robustness values than predictions of species extinctions. For example, for species extinctions on a real network \cite{kato2000} when true interaction strength values are used (as shown in Figure \ref{fig:inter_strength} \textbf{(a)}) and $T = 0.7$, we have $R_{species} = 0.619$. By contrast, for interaction loss on the same network when $T = 0.7$, we have $R_{inter} = 0.653$. Therefore, modelling secondary species extinctions as an outcome of interaction loss as opposed to primary species extinctions gives a different perspective on network robustness, allowing one to identify networks which are fragile against primary species loss but robust against interaction loss, and vice versa. A similar logic to that presented in Section \ref{sec:rand_extinct} may be followed in order to show that a network with a set number of primary species, secondary species and total interaction strength is maximally robust against interaction loss when the variance in interaction strength per secondary species is minimised.

\section{Discussion}

In conclusion, we have successfully extended the robustness framework of Memmott et al. \cite{Memmott2004} such that we may make predictions of ecological network robustness analytically. For random extinctions, we have shown that networks with low second species degree variance are highly robust. We are also able to predict secondary extinctions as primary species go extinct according to some degree or extinction risk based targeting, and we can predict secondary extinctions on ecological networks with more than two layers. Additionally, we can model the influence of random false positives and negatives in network data on robustness, finding that in small quantities false positives have a greater impact than false negatives on network robustness.

Our model is also capable of predicting the robustness of networks where interaction strength is weighted unevenly between different secondary species' neighbours, and species go extinct once a certain proportion of interaction strength has been lost. We have given results for robustness when interaction strength is equally distributed (maximally generalist), and when interaction strength is shared solely with one neighbouring species (maximally specialist). From this, we know that maximal generalisation and maximal specialisation can both produce a more robust network, depending on the proportion of interaction strength that must be lost before secondary species extinction. Finally, we have demonstrated the fact that it is also possible to model interaction strength loss as opposed to simply species extinction, representing a more ``gradual'' extinction scenario.

These results represent a substantial advancement in analytic understanding of ecological network robustness. However, there are still many open questions. We can predict the average secondary species extinction probability for a given number of primary extinctions, but we may also want to analytically predict the possible error in robustness curves by finding the standard deviation in secondary species extinction probability for a given number of primary extinctions. Additionally, we may want to establish the true worst and best case scenarios for secondary extinctions, as these have not been definitively identified. For errors (i.e. false positives and negatives) in network data, our current results examine errors which occur at random, but this may not be the case in the real world. Errors may occur due to some specific reason or dynamic, and identifying what this is may allow us to better mathematically model data errors and their influence.

Beyond these possible improvements, it is also important to acknowledge that in recent years, ecologists have considered properties of ecological networks which affect robustness and go beyond simpler models of species extinction. For example, ontogenetic niche shifts, where species change their diets when they undergo changes such as growing from a larvae to an adult, can affect the structure and robustness of interaction networks \cite{nakazawa2015}. Another consideration is how interactions can be ``rewired'' after species extinctions \cite{Schleuning2016,baldock2019}, which to predict analytically would likely require combinatoric methods for sampling with fuzzy replacement \cite{Kesemen2021}. These examples highlight the fact that there is still considerable room for analytic models of ecological network robustness to develop, and there are ongoing areas of research in both ecological networks and combinatorics which may complement one another well, so it may be useful for there to be a greater dialogue between these fields in future.

\section*{Funding}

C.J. was supported by Engineering and Physical Sciences Research Council Doctoral Training Partnership funding (EP/R513179/1). We thank the Research Focus Data-centric Sciences of the University of Potsdam for financial support.

\section*{Acknowledgements}

We thank Jane Memmott for providing feedback on a summary version of this paper.

\section*{Conflict of Interest Statement}

There authors have no conflict of interest to declare.

\section*{Data Availability}

The code used in generating the results for this paper may be found online at https://github.com/cj14373/eco-analytic.git

\bibliographystyle{ieeetr.bst}
\bibliography{biblio.bib}{}

\end{document}


\maketitle

\section*{False Positive and Negative Robustness Data}

In this section, we provide data for 18 different plant pollinator networks where we have measured their robustness against pollinator extinction ($R_{real}$) as well as their robustness when they have 20\% randomly added false positive edges ($R_{pos}$) and 20\% randomly removed false negative edges ($R_{neg}$). To determine whether false positives or false negatives have a larger impact on network robustness, we measure the net change in network robustness. This is calculated as $\frac{R_{pos} + R_{neg} - 2R_{real}}{R_{real}}$, so this measures net change relative to $R_{real}$. If the net change is positive, this means that false positives change robustness more than false negatives, and if the net change is negative, then false negatives have more impact than false positives. This data relates to results stated in Section 6.

\begin{longtable}{|l|l|l|l|l|l|} 
\hline
Network Name & $R_{real}$  & $R_{pos}$ & $R_{neg}$ & $\frac{R_{pos} + R_{neg} -   2R_{real}}{R_{real}}$ & References \\ \hline 
\endfirsthead
\endhead
MPL004       & 0.8246     & 0.8945    & 0.8035    & 0.0591                                            & \cite{barrett1987reproductive}  \\ \hline
MPL005       & 0.7770     & 0.8509    & 0.7602    & 0.0736                                            & \cite{clements1923experimental} \\ \hline
MPL009       & 0.8545     & 0.8925    & 0.8313    & 0.0174                                            & \cite{elberling1999structure}   \\ \hline
MPL015       & 0.9180     & 0.9999    & 0.9033    & 0.0733                                            & \cite{petanidou1991pollination} \\ \hline
MPL016       & 0.8760     & 0.9217    & 0.8625    & 0.0368                                            & \cite{herrera1988pollination}   \\ \hline
MPL021       & 0.7600     & 0.8644    & 0.7554    & 0.1312                                            & \cite{kato1990insect}           \\ \hline
MPL028       & 0.8089     & 0.8673    & 0.7965    & 0.0568                                            & \cite{primack1983insect}        \\ \hline
MPL029       & 0.8142     & 0.8556    & 0.7894    & 0.0204                                            & \cite{primack1983insect}        \\ \hline
MPL043       & 0.8335     & 0.8792    & 0.8163    & 0.0343                                            & \cite{montero2005ecology}       \\ \hline
MPL044       & 0.7680     & 0.8514    & 0.7589    & 0.0969                                            & \cite{kato2000}                 \\ \hline
MPL047       & 0.8827     & 0.9371    & 0.8683    & 0.0454                                            & \cite{dupont2009ecological}     \\ \hline
MPL048       & 0.8583     & 0.9316    & 0.8514    & 0.0775                                            & \cite{dupont2009ecological}     \\ \hline
MPL049       & 0.8846     & 0.9250    & 0.8704    & 0.0297                                            & \cite{bek2006pollination}       \\ \hline
MPL054       & 0.7473     & 0.8146    & 0.7303    & 0.0674                                            & \cite{kakutani1990insect}       \\ \hline
MPL055       & 0.7457     & 0.8143    & 0.7343    & 0.0768                                            & \cite{kato1996flowering}        \\ \hline
MPL056       & 0.8125     & 0.8694    & 0.7960    & 0.0496                                            & \cite{kato1993flowering}        \\ \hline
MPL057       & 0.8020     & 0.8945    & 0.7906    & 0.1011                                            & \cite{inoue1990insect}          \\ \hline
MPL062       & 0.9695     & 1.0000    & 0.9628    & 0.0245                                            & \cite{Robertson1929}            \\ \hline
\caption[Real world ecological network data for robustness with and without false positives and negatives]{Data table of 18 real world plant pollinator networks, where datasets were accessed from \cite{fortuna2014web}. Data regarding network names, real robustness, robustness values with 20\% false positives and negatives, net changes in robustness and references are included.}
\label{tbl:eco_pos_neg}
\end{longtable}

In Table \ref{tbl:eco_pos_neg}, we can see that for all networks the net change in robustness is positive. This indicates the fact that false positives have a greater impact on network robustness than false negatives. The average net change is $5.95\%$, and the median net change is $5.80\%$.

\section*{Deterministically Estimating the Weight Sum}

Here we present the details behind our algorithm for estimating values of $Pr(\sum_{0}^j W \geq i_A)$, as discussed in Section 7. Given some weight vector $\textbf{w}$ of length $k_A$ with elements $w_z$ for some species $A$, we want to know how many combinations of $j$ elements of $\textbf{w}$ sum up to $\geq i_A$, the extinction threshold for species $A$. We may express our choices of weights with $\textbf{y}$, a vector with elements $y_z$ which can take values of 0 or 1, and $\sum_{z=0}^{k_A} y_z = j$. Values of $\textbf{y}$ correspond to whether or not an element of $\textbf{w}$ has been chosen, so if $y_z = 1$ then the $z^{th}$ element of $\textbf{w}$ has been chosen. We may then calculate $\sum_{0}^j W$ as $\sum_{z=0}^{k_A} w_z y_z = \textbf{w} \cdot \textbf{y}$. There are a total of ${k_A \choose j}$ possible combinations of weights, so it is computationally impractical to calculate all possible combinations of weights above small values of $k_A$ and $j$. Instead, it is necessary to estimate the number of weight combinations which meet or exceed the threshold $i_A$ in order to calculate $Pr(\sum_{0}^j W \geq i_A)$.

As mentioned in the main text, it is possible to express $\textbf{y}$ as a binary number and provide an ordering of possible values of $\textbf{y}$ such that, were $w_z$ values powers of $2$ ordered in descending value (i.e. $w_z = 2^{k_A-z}$), it would always be possible to identify the $n^{th}$ ordering of $y$ above which $\textbf{w} \cdot \textbf{y} \geq i_A$ and below which all $\textbf{w} \cdot \textbf{y} < i_A$. However, this is not necessarily a realistic sequence of $w_z$ weights. In Table \ref{tbl:weight_sums}, we provide two example $\textbf{w}$ vectors of length $k_A = 4$, where weights are in descending size order. In the $\textbf{y}$ vector column, we give each possible ordering of $\textbf{y}$ for which $\sum_{z=0}^{k_A} y_z = 2$. The $\textbf{y}$ vectors are themselves ordered such that they are in ascending size order for the binary numbers the correspond to (e.g. $\textbf{y} = (0,1,0,1)$ corresponds to the binary number $0101$, which is $5$ in base 10).

\begin{longtable}{|c|c|c|}
\hline
$\textbf{y}$ & \begin{tabular}[c]{@{}c@{}}$\textbf{w} \cdot \textbf{y}$\\ $\textbf{w} = (8,4,2,1)$\end{tabular} & \begin{tabular}[c]{@{}c@{}}$\textbf{w} \cdot \textbf{y}$\\ $\textbf{w} = (3,3,2,1)$\end{tabular} \\ \hline
\endfirsthead
\endhead
(0,0,1,1)    & 3                                                                                            & 3                                                                                            \\ \hline
(0,1,0,1)    & 5                                                                                            & 4                                                                                            \\ \hline
(0,1,1,0)    & 6                                                                                            & 5                                                                                            \\ \hline
(1,0,0,1)    & 9                                                                                            & 4                                                                                            \\ \hline
(1,0,1,0)    & 10                                                                                           & 5                                                                                            \\ \hline
(1,1,0,0)    & 12                                                                                           & 6                                                                                            \\ \hline
\caption[Ordering of $\textbf{y}$ vector and weight sums]{Ordering of $\textbf{y}$ vector with corresponding weight sum values for different $\textbf{w}$ vectors.} \label{tbl:weight_sums}
\end{longtable}

 We can see that when $\textbf{w} = (8,4,2,1)$, the values of $\textbf{w} \cdot \textbf{y}$ are in ascending size order, and so it is straightforward to identify the ordering of $\textbf{y}$ above which $\textbf{w} \cdot \textbf{y} \geq i_A$ and below which $\textbf{w} \cdot \textbf{y} < i_A$ for any given $i_A$. However, when $\textbf{w} = (3,3,2,1)$ the values of $\textbf{w} \cdot \textbf{y}$ are not ordered, and so if we were to set $i_A = 5$ for example, we cannot identify an ordering of $\textbf{y}$ above which $\textbf{w} \cdot \textbf{y} \geq i_A$ and below which $\textbf{w} \cdot \textbf{y} < i_A$.

 While in this example it is straightforward to explicitly enumerate all possible orderings of $\textbf{y}$, this is impractical for for larger $k_A$ and $j$ values. Therefore, we must consider how to estimate the number of orderings of $\textbf{y}$ for which $\textbf{w} \cdot \textbf{y}$. If we order values of $\textbf{w}$ in descending size order, then it is possible to establish certain groups of $\textbf{y}$ vectors which will always give $\textbf{w} \cdot \textbf{y} \geq i_A$ or $\textbf{w} \cdot \textbf{y} < i_A$. For example, let us consider the $\textbf{w} = (3,3,2,1)$ and $\sum_{z=0}^{k_A} y_z = 2$ scenario from Table \ref{tbl:weight_sums}. If we set the threshold $i_A = 4$, then we know that all $\textbf{y}$ vectors of the forms $(1,...)$ and $(0,1,...)$ must give $\textbf{w} \cdot \textbf{y} \geq i_A$, and all $\textbf{y}$ vectors of the form $(0,0,...)$ must give $\textbf{w} \cdot \textbf{y} < i_A$. Therefore, for this example we would only need to sample 3 orderings of $\textbf{y}$ in order to accurately estimate the value of $Pr(\sum_{0}^j W \geq i_A)$, as opposed to having to sample all 6 orderings.

 We can extend this idea in order to develop an estimation algorithm. Given some $\textbf{w}$ weights arranged in descending size order, we specify some prefix of $\textbf{y}$ (referred to as $\textbf{y}^{p}$) which gives the first $z = 1,2,...,x$ values of $\textbf{y}$. We then create a ``bracket'' for the prefix $\textbf{y}^{p}$, finding $\textbf{y}^{p}_{upper}$ and $\textbf{y}^{p}_{lower}$ which are the highest and lowest orderings of $\textbf{y}$ with prefix $\textbf{y}^{p}$ respectively. If $\textbf{y}^{p}_{upper}$ and $\textbf{y}^{p}_{lower}$ satisfy $\textbf{w} \cdot \textbf{y}^p \geq i_A$, then we count all of the orderings in the bracket towards our estimation of $Pr(\sum_{0}^j W \geq i_A)$. If $\textbf{y}^{p}_{upper}$ and $\textbf{y}^{p}_{lower}$ fulfil $\textbf{w} \cdot \textbf{y}^p < i_A$, then we simply discount the prefix. If $\textbf{w} \cdot \textbf{y}^p_{upper} \geq i_A$ and $\textbf{w} \cdot \textbf{y}^p_{lower} < i_A$, then we lengthen the prefix and repeat. We can specify some maximum depth $d$ of prefix, where $\sum_{z=0}^{k_A} y^{p}_{z} \leq d$, i.e. $d$ is the maximum number of $y_z = 1$ values permitted in the prefix. This allows us to perform the estimation while specifying an endpoint, preventing us from searching through too many orderings of $\textbf{y}^{p}$ and conserving computational resources.

 If we reach the maximum prefix depth such that $\sum_{z=0}^{k_A} y^{p}_{z} = d$ and we still have $\textbf{w} \cdot \textbf{y}^p_{upper} \geq i_A$ and $\textbf{w} \cdot \textbf{y}^p_{lower} < i_A$, then we estimate the number of orderings in the bracket which give $\textbf{w} \cdot \textbf{y} \geq i_A$. If we set $\textbf{w} \cdot \textbf{y}^p_{upper} = U$ and $\textbf{w} \cdot \textbf{y}^p_{lower} = L$, and the number of orderings in the bracket $b$ as $s_b$, then we estimate the number of orderings $n_b$ in the bracket for which $\textbf{w} \cdot \textbf{y} \geq i_A$ as

 \begin{equation}
     n_b \approx \Big\lfloor \frac{U - i_{A} + 1}{U - L} s_b \Big\rfloor,
 \end{equation}

 which is accurate if the values of $\textbf{w} \cdot \textbf{y}$ are equally spaced in the bracket. We express this algorithm as Algorithm \ref{alg:weight} in pseudocode form below.

 \begin{algorithm}
 \begin{algorithmic}[1]
\caption{Deterministic Weight Sum Algorithm}\label{alg:weight}
\While{$\sum_{z=0}^{k_A} y^{p}_{z} > 0$ and $\sum_{z=0}^{k_A} y^{p}_{z} \leq j$ and $k_A - \text{length}(\textbf{y}^p) \geq j - \sum_{z=0}^{k_A} y^{p}_{z}$}
\State Specify some prefix $\textbf{y}^{p}$ and calculate corresponding $U$ and $L$ values
\If{$U \geq i_A$ and $L \geq i_A$}
\State Record $n_b$ = $s_b$
\State Remove all values from $\textbf{y}^{p}$ after and including the last non-zero value of $\textbf{y}^{p}$
\State Append (0,1) to $\textbf{y}^{p}$
\ElsIf{$U < i_A$ and $L < i_A$}
\If{$\sum_{z=0}^{k_A} y^{p}_{z} \geq 2$}
\State Remove all values from $\textbf{y}^{p}$ after and including the second to last non-zero value of $\textbf{y}^{p}$
\State Append (0,1) to $\textbf{y}^{p}$
\ElsIf{$\sum_{z=0}^{k_A} y^{p}_{z} < 2$}
\State Remove all values from $\textbf{y}^{p}$
\EndIf
\ElsIf{$U \geq i_A$ and $L < i_A$}
\If{$\sum_{z=0}^{k_A} y^{p}_{z} \leq d$}
\State Append (1) to $\textbf{y}^{p}$
\ElsIf{$\sum_{z=0}^{k_A} y^{p}_{z} > d$}
\State Record $n_b = \Big\lfloor \frac{U - i_{A} + 1}{U - L} s_b \Big\rfloor$
\State Remove all values from $\textbf{y}^{p}$ after and including the last non-zero value of $\textbf{y}^{p}$
\State Append (0,1) to $\textbf{y}^{p}$
\EndIf
\EndIf
\EndWhile
\State \Return{$Pr(\sum_{0}^j W \geq i_A) = \frac{\sum_b n_b}{{k_A \choose j}}$}
\end{algorithmic}
\end{algorithm}

\begin{figure}[h]
    \begin{minipage}{1\textwidth}
    \centering
    \makebox[\textwidth]{\includegraphics[width = 1\textwidth]{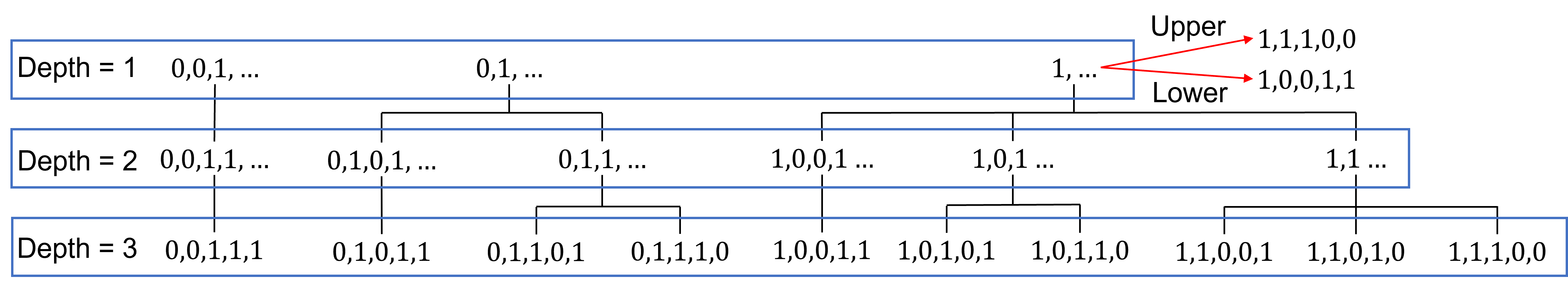}}
    \vspace{-10pt}
    \caption[Dendrogram illustrating different $\textbf{y}^p$ prefixes at increasing levels of depth]{Dendrogram illustrating different $\textbf{y}^p$ prefixes at increasing levels of depth. For depth = 1, we show that for $\textbf{y}^p = (1)$, we get $\textbf{y}^{p}_{upper} = (1,1,1,0,0)$ and $\textbf{y}^{p}_{lower} = (1,0,0,1,1)$. As we increase the depth, we can see that we specify more prefixes.}
    \label{fig:binary_dendro}
    \end{minipage}
\end{figure}

In order to illustrate the importance of the depth parameter in providing an accurate result for $Pr(\sum_{0}^j W \geq i_A)$, we provide an example in the following. Let us consider a system where $\textbf{w} = (5,5,3,2,1)$, $j = 3$ and $i_A = 9$. In Figure \ref{fig:binary_dendro}, we provide a dendrogram representation of the different possible $\textbf{y}^p$ prefixes, sorted vertically by depth (i.e. value of $\sum_{z=0}^{k_A} y^{p}_{z}$) and horizontally by binary number value represented by $\textbf{y}$.

For the prefix $\textbf{y}^p = (1)$, we get $\textbf{y}^{p}_{upper} = (1,1,1,0,0)$ and $\textbf{y}^{p}_{lower} = (1,0,0,1,1)$, resulting in $U = 13$ and $L = 7$ respectively. For this prefix at depth 1, $U > i_A$ and $L < i_A$. If we set the maximum depth $d = 1$, then we need to estimate $n_b$ as $\Big\lfloor \frac{U - i_{A} + 1}{U - L} s_b \Big\rfloor = 4$, whereas the real value for $n_b$ is actually 5. If we run our algorithm with the conditions $\textbf{w} = (5,5,3,2,1)$, $j = 3$, $i_A = 9$ and $d = 1$, we get an estimation of $Pr(\sum_{0}^j W \geq i_A) = 0.5$, whereas the true probability is $0.7$. Once we increase the depth to $d = 2$, then we get and estimation of $Pr(\sum_{0}^j W \geq i_A) = 0.7$, but this comes at the cost of evaluating more prefixes and therefore greater computational expense. As we show in the main text, this method is more accurate than Monte Carlo methods when run for comparable lengths of time. Additionally, it is deterministic, so for a given set of intial parameters it will always return the same estimate of $Pr(\sum_{0}^j W \geq i_A)$.

\bibliographystyle{ieeetr.bst}
\bibliography{biblio_sup.bib}{}